# Enhancement of positive magnetoresistance following a magnetic-field-induced ferromagnetic transition in an intermetallic compound, $Tb_5Si_3$

S. Narayana Jammalamadaka, Niharika Mohapatra, Sitikantha D Das, and E.V. Sampathkumaran[*]

*Tata Institute of Fundamental Research, Homi Bhabha Road, Colaba, Mumbai 400005, India.*

We report the existence of a field-induced ferromagnetic transition in the magnetically ordered state (<69 K) of an intermetallic compound, $Tb_5Si_3$, and this transition is distinctly first-order at 1.8 K (near 60 kOe), whereas it appears to become second order near 20K. The finding we stress is that the electrical resistivity becomes suddenly large in the high-field state after this transition and this is observed in the entire temperature range in the magnetically ordered state. Such an enhancement of 'positive' magnetoresistance (below 100 kOe) at the metamagnetic transition field is unexpected on the basis that the application of magnetic-field should favor a low-resistive state due to alignment of spins.

It is well-known that, in metals containing magnetic moments, the electrical resistivity (ρ) gets reduced in the ferromagnetic and paramagnetic state with the application of a magnetic-field (H). This arises from the reduction of magnetic scattering following the suppression of magnetic fluctuations by the applied magnetic field. As a result, negative magnetoresistance [MR, defined as $\{\rho(H)-\rho(0)\}/\rho(0)$] in such systems is usually observed. In the antiferromagnetic cases, ferromagnetic alignment can be abruptly induced by an applied field. In such 'metamagnetic' cases as well, at the metamagnetic transition field ($H_t$), negative MR is expected as demonstrated for many rare-earth compounds [1, 2] in which indirect exchange interaction controls magnetism. Another family of materials in which negative MR at $H_t$ attracted attention in recent years is 'giant magnetoresistive manganites'[3] in which case double-exchange interaction plays a major role in mediating magnetic interaction.

In this Rapid Communication, we present evidence for the opposite behavior of MR at the metamagnetic transition in an intermetallic compound, $Tb_5Si_3$, known to form in $Mn_5Si_3$-type hexagonal structure (space group: $P6_3/mcm$) [4]. This compound has been known to undergo a complex helimagnetic ordering below ($T_N=$) 69 K [5, 6]. We observe a field-induced ferromagnetic transition in the magnetically ordered state. This transition is discontinuous in its character at 1.8 K (near 60 kOe), but continuous at higher temperatures and $H_t$ systematically decreases with increasing temperature. The key experimental finding is that, despite these differences at various temperatures, in the magnetically ordered state, ρ is dramatically enhanced at all temperatures resulting in large positive MR at the metamagnetic transition compared to the values at low fields (below $H_t$), in contrast to expectations. The positive sign of MR is interestingly retained in the entire field range of investigation (<100 kOe).

The sample in the polycrystalline form was prepared by arc melting together stoichiometric amounts of high-purity Tb (>99.9 wt%) and Si (>99.99 wt%) in an atmosphere of argon. The sample thus prepared was found to be single phase within the detection limit (<2%) of x-ray diffraction (Cu $K_\alpha$). We could obtain a very good Rietveld fitting of the x-ray diffraction pattern assuming the occupation of Tb and Si at respective sites and the lattice constants (± 0.004 Å) derived from such a fitting are: $a=$ 8.441 and $c=$ 6.347 Å. Absence of any extra phase was also confirmed by back-scatterred scanning electron microscopic image and the uniformity of the composition was further confirmed by energy dispersive x-ray analysis. Dc magnetization (M) measurements down to 1.8 K were performed with the help of a commercial (Oxford Instruments) vibrating sample magnetometer. The electrical resistance measurements (1.8 – 300 K) were performed in the presence of magnetic fields by a commercial (Quantum Design) Physical Property Measurements System (PPMS) by a conventional four-probe method in the transverse geometry, that is, with the electric current (I) perpendicular to field-direction (unless otherwise stated). The electrical contacts of the leads to the specimen were made by a conducting silver paste. In order to identify magnetic transitions, we have also performed heat-capacity (C) measurements (1.8 – 250 K) with the same PPMS by the relaxation method and also ac susceptibility (χ) measurements ($H_{ac} = 1$ Oe) with several frequencies (1.3 to 1333 Hz).

We show the results of measurements of *ac* χ, *C* and *ρ* in figure 1 below 150 K [7], essentially to confirm magnetic transition temperature. All these measurements were performed after zero-field-cooling (ZFC) of the specimen from 150 K (from the paramagnetic state). As seen in figure 1, there is a peak in *ac* χ(T) near 70 K; below 69 K, there is a sharp decrease of χ. Similar features were seen in the *dc* χ(T) measured with *H=* 5 kOe and therefore not shown here. The fact that the magnetic ordering does not arise from spin-glass freezing is evidenced by the fact that the peak in real part of *ac* χ is found to be frequency-independent (figure 1*a*). No feature was observable in the imaginary part of the *ac* χ, which demonstrates that this magnetic feature is due to local moments. There is an upturn of *C* below 69 K with a well-defined peak in *C(T)* at 67 K (see figure 1*b*); this peak is quite sharp and narrow, endorsing the absence of any type of disordered magnetism at this temperature. With respect to $\rho$, the absolute values may not be reliable due to micro-cracks in the polycrystalline sample as well as due to the spread of silver-paint. Hence we show the normalized plot in figure 1*c*. The value of ρ at 300 K is typically of the order of a few hundred μΩcm and the residual resistivity ratio (~ 2.5) is nearly the same as that observed in Ref. 6. In zero field, there is a sharp fall of *ρ* below 69 K due to the loss of spin-disorder contribution (see figure 1*c*) as known earlier [6]. The main point of emphasis is that a magnetic transition takes place near 69 K.

In order to bring out the existence of a metamagnetic transition in the magnetically ordered state, we show isothermal *M* behavior in figure 2. The data were collected (for the ZFC condition of the specimen) while sweeping the magnetic-field at the rate of 4 kOe/min in very close intervals of *H*. Looking at the *M(H)* plot at 1.8 K, *M* increases gradually with *H* and, at ($H_t=$) 58.5 kOe, there is a sharp increase indicating field-induced ferromagnetic alignment. Beyond this field, the variation is quite sluggish. As the field is reversed, the curve follows the virgin curve closely till about 65 kOe below which the gradual fall of *M* continues till about ($H_t^r=$) 27 kOe (deviating from the path of the virgin curve) at which there is a sharp fall merging with the virgin curve. Thus, there is a field-induced first-order magnetic transition which is hysteretic in the intermediate field range. When the field-direction is reversed, we see exactly the same features (not shown in the figure), unlike the situation in $Nd_7Rh_3$ (Ref. 2). As the temperature is raised, say, to 5 K, similar features appear near the same fields, except for the fact that the metamagnetic transition is broadened. At 20 K, the field-induced transition appears at a lower field (with the mid-point of the transition near 50 kOe) in the forward cycle, with a significantly reduced hysteresis loop. Thus, it appears that the field-induced transformation becomes second order in the vicinity of 20 K ("a tricritical point"). For 50 K, the loop is almost absent with the broadened transition occurring near 39 kOe. Thus, the transition appears at lower fields with increasing temperature. The trends observed in the transition fields and hysteresis loop with increasing temperature are consistent with the expectations for first-order magnetic transitions [8]. Before completion of this part of discussion, we point out that that the magnetic moment, say at 120 kOe, is about 8 $\mu_B$/Tb (in the magnetically ordered state) which is very close to the free ion value of 9 $\mu_B$/Tb. This endorses the conclusion that the compound attains ferromagnetism at such high fields. Needless to state that, above 70 K, we did not observe any such field-induced transitions.

We now present the magnetoresistance behavior to focus on the point of central emphasis. The results are shown in detail in figure 3 at various temperatures. Before collecting the data as a function of *H* (0 to 100 to -100 to 100 to 0 kOe) for each temperature, the sample was warmed up to paramagnetic state and cooled in zero-field to the desired temperature, as we noticed a small bifurcation of χ below $T_N$ for the ZFC and field-cooled conditions of measurements. Looking at the *MR* curve for T= 1.8 K, *MR* is positive, and this sign for H below metamagnetic transition is expected for antiferromagnetism (without magnetic Brillioun-zone gap). The usual Lorentz force contribution in metals also results in positive sign of MR with a negligibly small value. The most fascinating finding is that, at $H_t$, there is a sudden increase in *ρ* and the magnitude of *MR* increases from less than 1% (below $H_t$) to about 40% at 63 kOe (see virgin curve in figure 3). This enhancement of *ρ* relative to low-field state is in sharp contrast to what one would have expected for a transformation to ferromagnetic alignment as stated earlier. Further increase of *H* after crossing the transition gradually reduces [9] the magnitude of *MR*, resulting in negative slope of *d(MR)dH* typical of a ferromagnet. On reducing the field after reaching the highest field (100 kOe), the curve traces back the virgin curve till $H_t$. However, it is also interesting to note that, below $H_t$, the value of $\rho$ keeps rising with decreasing field till $H_t^r$ without tracking the virgin curve. Precise origin of this intriguing finding is not clear at present, though it is possible that this is a hysteresis effect with the ferromagnetically aligned specimen showing up its dominating electrical transport in the intermediate field range. Near $H_t^r$, there is a sharp fall with the data essentially falling on the virgin curve. The value of $H_t^r$ in this cycle in *MR* data is marginally lower (19 kOe) compared to that in *M(H)* data. It appears that such differences do occur [10] if the modes of measurements (whether the data is collected after stabilizing the field or while sweeping the field) are different. Possibly strains arising from the way the field/temperature cycling is done contributes to such small differences. The value of *MR* near $H_t^r$ attains as large as about 160% (for the transverse geometry). Needless to emphasize that the *MR(H)* behavior described above is seen even when the magnetic field is reversed, unlike in $Nd_7Rh_3$ (Ref. 2). We also noted a weak irreversibility in $H_t^r$ value (about 27 kOe) after cycling through reverse field values, possibly due to strains. Finally, we have also measured *MR* in the longitudinal geometry (I // H) and, as shown in figure 3, the features are reproduceable, except for the fact that value of MR gets gradually higher in the reverse cycle below about 45 kOe in the intermediate field range attaining a value of about 180% at about 26 kOe before a sharp 'fall' and subsequent merger with the virgin curve.

    It is of interest to see how the *MR(H)* curve gets modified with varying temperature. It is important to note that, despite broadened metamagnetic transition, the key finding is distinctly preserved at 5 and 10 K, though the hysteresis-loop region gets narrower (see figure 3). Even at 20 K, at which the hysteretic loop is absent, the resistivity following spin-reorientation is higher. It is to be noted that, even near the magnetic ordering temperature, say at 60 K, the features are essentially the same as that at lower temperatures (comparing with the curve for 20 K). Onset of positive *MR* just below $T_N$ is obvious when one compares $\rho(T)$ curves of *H*= 0 and 50 kOe in figure 1*c*. Above 70 K (see the curves for 80 K and 110 K), the sign of *MR* is distinctly negative and the magnitude of *MR(H)* gradually decreases typical of paramagnets. Thus, the anomalous features evolve in the magnetically ordered state only [11].

    The above-described *MR* enhancement must have its origin in ferromagnetism, as it happens at the onset of metamagnetic transition. Therefore, any explanation should consider this fact. It is not clear whether there is a profound change in the Fermi surface at high-fields (>$H_t$), following metamagnetic transition. While such a Fermi surface change is possible in Ce systems due to some degree of itinerancy of 4f electrons as demonstrated recently [12], it is not expected for strictly 'localized-4f' metallic systems like that of Tb. Positive sign of *MR* in the ferromagnetic state has been reported [13] in geometrically constrained ferromagnetic metals [14]. It would be surprising if any geometrical confinement effect of electronic transport as proposed in Ref. 14 takes place in these bulk materials. Such an explanation is not applicable for the present case, as the 4f-electrons of Tb are strictly localized. Role of grain boundary effects as a possible cause of the observed anomaly is ruled out as it is well-known [15] that such an effect should result in negative MR. While it is not clear whether this compound presents an exceptional situation in which the Lorentz force contribution becomes 'giant' at the metamagnetic transition due to the internal magnetic fields generated due to the onset of ferromagnetic alignment, overcompensating the scattering contribution by Tb local spins, such

an explanation appears to be a contradiction if one compares MR behavior in transverse and longitudinal geometry. Generally speaking [16], for usual Lorentz force contributions, it is expected that the MR values in the transverse geometry are higher than those for longitudinal geometry. However, we find that the MR curves either overlap for both the configurations (in some portions of the plot) or lie higher in some field range for the latter configuration in the magnetically ordered state, in contrast to the above expectation. When internal fields aid Lorentz force, it is not clear how MR in these two geometries are expected to differ. At this juncture, we would like to mention that we have measured *M(H)* and *MR(H)* of other heavy rare-earths (Gd, Dy and Ho) in the same family. Though no sharp metamagnetic transitions were observed in these compounds, there is a broad spin-reorientation transition in M(H) around 40 and 20 kOe for Gd and Ho respectively at 1.8 K. However, we do not find any anomaly in the sign of MR at the spin-reorientation fields. One would have naively expected that an argument in terms of an anomalous Lorentz force contribution as discussed above for $Tb_5Si_3$ yields an 'enhanced' *positive* MR at high fields in these compounds as well, in contrast to the observation. Thus, the MR behavior of this compound is a puzzle. Whatever be the microscopic origin, the finding – enhanced ρ *following field-induced first-order metamagnetic transition* - for $Tb_5Si_3$ alone in this family presents an interesting situation, though the magnitude of 'positive' MR gradually increasing with H to large values have been known in some non-magnetic systems [17].

To conclude, we have identified a 'local-moment' compound, in which the electrical resistivity of the field-induced ferromagnetic state gets dramatically higher at the onset of metamagnetic transition. The sign of the MR remains positive with a *large magnitude* following this transition. It is also intriguing to note that even in an intermediate-field range, while reversing the magnetic-field direction, this 'relatively more resistive' ferromagnetic state apparently dominates the scattering process. This work demonstrates that there are still unexplored puzzles in the magnetoresistance behavior of magnetic metals, particularly across metamagnetic transitions. [Finally, this investigation on $Tb_5Si_3$ was primarily motivated to make sure that the anomalous metamagnetic transitions observed in $Tb_6Co_{1.67}Si_3$ by us recently (see Ref.10) is not due to the impurity $Tb_5Si_3$.]

We thank Kartik K Iyer for his help during the measurements.

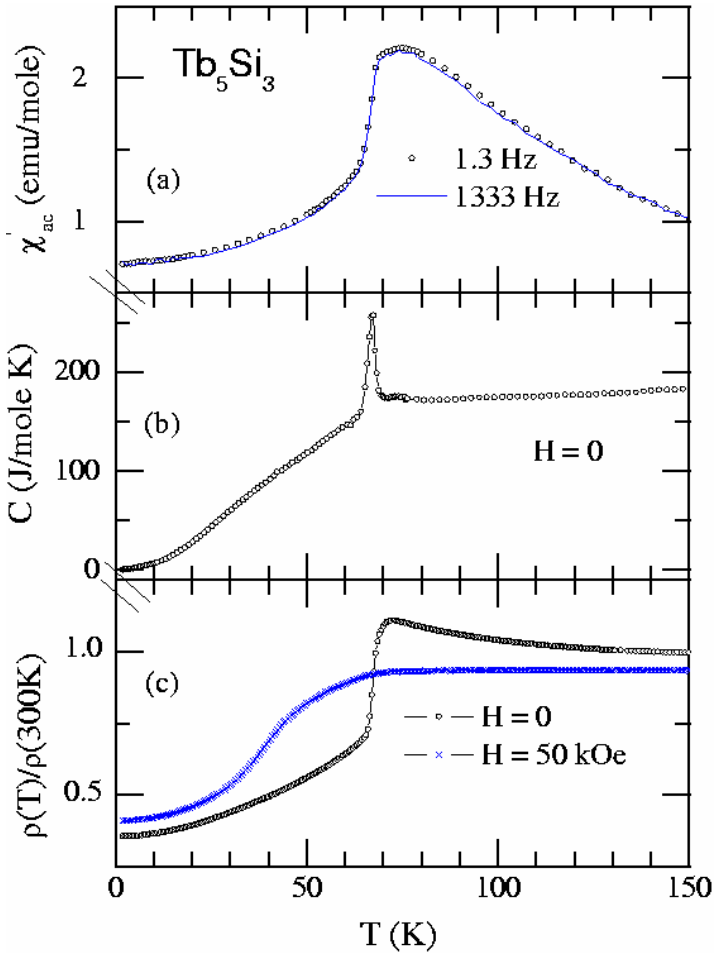

Figure 1:
(color online) **(a)** *R*eal part of *ac* $\chi$, **(b)** heat-capacity (*C*), and **(c)** normalized electrical resistivity ($\rho$) as a function of temperature for $Tb_5Si_3$. We have also plotted the $\rho$ data in the presence of 50 kOe. In the case of *C(T)* figure, a line is drawn through the data points.

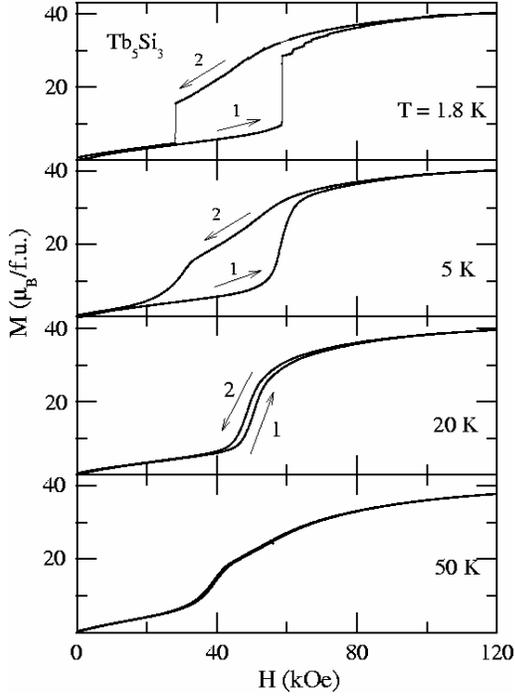

Figure 2:
Isothermal magnetization behavior at several temperatures for $Tb_5Si_3$. The numericals near the curves serve as guides to the eyes.

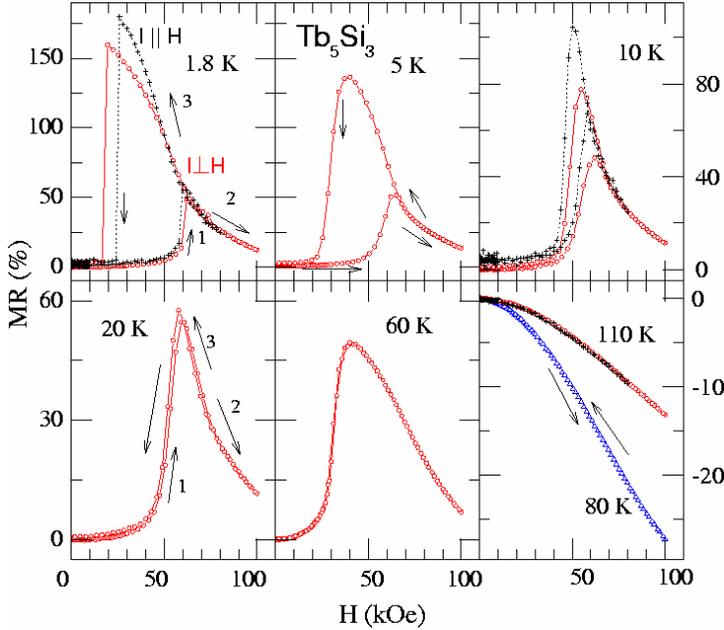

Figure 3:
(color online) Magnetoresistance, defined as $[\rho(H)-\rho(0)]/\rho(0)$, as a function of magnetic field for $Tb_5Si_3$ for transverse geometry of current and field directions (data points with continuous lines through them). For some temperatures, the curves (dotted lines through points marked '+') are shown for longitudinal geometry. The arrows and numbers placed near these (shown in some graphs only) serve as guides to the eyes.